\documentclass{JHEP3}

\usepackage[dvips]{graphicx}
\usepackage{bm}
\usepackage{amsmath,amssymb}
\usepackage{comment}

\newcommand{\nt}{\nonumber\\}

\newcommand{\cF}{{\cal F}}

\newcommand{\cS}{{\cal S}}
\newcommand{\cT}{{\cal T}}

\newcommand{\vR}{{\vec R}}

\newcommand{\ba}{\begin{eqnarray}}
\newcommand{\ea}{\end{eqnarray}}

\newcommand{\eps}{\epsilon}

\newcommand{\tls}{\tilde s}
\newcommand{\tlt}{\tilde t}
\newcommand{\tlr}{\tilde r}

\input{colordvi.tex}

\title{Microstates of black holes in expanding universe from interacting branes}

\preprint{KEK-TH-1943} 

\author{
Shotaro Shiba
\vspace*{3mm}\\
Theory Center, High Energy Accelerator Research Organization (KEK),\\
1-1 Oho, Tsukuba, Ibaraki 305-0801, Japan.
\vspace*{3mm}\\
\email{sshiba@post.kek.jp}
}

\abstract{
Thermodynamics of the near extremal black $p$-branes can be described by collective 
motions of gravitationally interacting branes.
This proposal is called the $p$-soup model.
In this paper, we check this proposal in the case of black brane system which is 
asymptotically Friedmann-Lema\^i{}tre-Robertson-Walker universe in an infinite distance.
As a result, we can show that the gravitationally interacting branes explain 
free energy, entropy, temperature and other physical quantities in these systems.
This implies that the microstates of this kind of 
brane system can be also understood in the $p$-soup model.
}

\begin{document}

\section{Introduction}


Microstates of black holes are still an outstanding problem in theoretical physics. 
This discussion was initiated by 
Strominger and Vafa~\cite{Strominger:1996sh}.
In their picture, branes are static in noncompact spacetime 
and strings on the branes provide dynamical degrees of freedom. 
Such studies have been mainly developed in the intersecting black branes, 
especially in the D1-D5 system~\cite{Callan:1996dv}. 
(See also a review \cite{David:2002wn}.)
However, it is still unclear whether they can be generalized to other various types 
of black holes. 

Recently we proposed another description of the black hole microstates~\cite{Wiseman:2013cda, Morita:2013wfa, Morita:2014ypa}.
In our picture, branes are moving at the speed proportional to Hawking temperature.
They have kinetic energy and strongly gravitationally interacting with each other,
then compose a bound state at low energy.
We can regard this bound state as a black brane.

More concretely, we impose the following settings for the system: 
\begin{itemize}
\item The characteristic velocity $v$ of the branes should satisfy the condition 
$v \propto \pi \cT r$,
where $r$ is the characteristic size of the system and $\cT$ is Hawking temperature.

\item If the effective action for the branes is expanded in a series of gravitational coupling, all the terms should be of the same order.
\end{itemize}
The first setting can be understood as the condition which 
Matsubara modes satisfy in systems at finite temperature.
The second setting is a kind of virial theorem for systems with strong gravitational coupling.
We call this proposal the $p$-soup model.


In the previous papers, using these settings, 
we discussed the systems of parallel D- or M-branes~\cite{Morita:2013wla}
and intersecting D- or M-branes~\cite{Morita:2014cfa, Morita:2015jga}.
We analyzed these systems in our picture, 
then we could correctly estimate free energy, horizon size and other physical quantities.
The results are consistent with those of the corresponding black branes.

Here we note that while the branes are moving in our picture,
these corresponding black branes are static solutions in supergravity.
In some cases, momentum can be introduced in an isometry direction,
but the time dependence is so limited. 

In this paper, we discuss more nontrivial time-dependent systems.
A time-dependent solution in supergravity can be obtained as a simple generalization 
of a static black branes~\cite{Maeda:2009zi}.
Therefore, we analyze the microstates of such solutions in the $p$-soup model, 
and check if our discussion can be applicable to the time-dependent brane systems.



As a result, we can successfully show that gravitationally interacting branes explain
correct physical quantities of the time-dependent black branes.
Compared with the static black branes, unfortunately,
some uncertainty appears.
Including such a subtle point, we discuss these systems in detail.


This paper is organized as follows.
In \S \ref{sec2}, we review the time-dependent black brane solutions in Einstein-Maxwell-dilaton theory.
In \S \ref{sec3}, we analyze such black brane systems based on the $p$-soup model
and check if we can reproduce the supergravity results (summarized in appendix \ref{secA}).
In \S \ref{sec4}, we conclude our discussion.

\section{Time-dependent black brane solution}
\label{sec2}

We consider $D$-dimensional gravitational theory coupled to dilaton $\phi$ and $(n_A+1)$-form field.
The action is
\ba \label{SD}
S_D=\frac{1}{16\pi G_D} \int d^Dx \sqrt{-g}\left[
R-\frac12(\partial\phi)^2-\sum_A\frac{1}{2(n_A+2)!} e^{a_A\phi} F_{n_A+2}^2
\right]
\ea
where we set
\ba
a_A^2 = 4-\frac{2(n_A+1)(D-n_A-3)}{D-2}\,,
\ea
so that we have asymptotically flat spacetime solutions.

The solutions are 
understood as intersecting brane systems.
In the extremal limit, the metric can be written as
\ba \label{met-D}
ds_D^2 
= \prod_A H_A^{\frac{q_A+1}{D-2}} \left[
-\prod_A H_A^{-1} dt^2
+\sum_{\alpha=1}^{D-d} \prod_A H_A^{-\delta_A^{(\alpha)}} dy_\alpha^2
+\sum_{i=1}^{d-1} dx_i^2
\right]
\ea
where $H_A$ is a harmonic function in 
$(d-1)$ dimensions $x_i$.
$q_A$ is the spatial dimension of brane $A$.
$\delta_A^{(\alpha)}$ equals 1 if the brane $A$ is expanded in the direction $y_\alpha$,
and otherwise $\delta_A^{(\alpha)} = 0$.
The index $A$ denotes species of branes.\footnote{
If branes of the same species are expanded in different directions,
we distinguish them here.}
The harmonic functions $H_A$ are usually time-independent:
\ba
H_{A_S} = 1+\frac{Q_{A_S}}{r^{d-3}}\,.
\ea
However, we can generalize them by making some of them time-dependent\footnote{
If there are more than one species of time-dependent branes ($n_T>1$),
we need an additional potential for the dilaton $\sim e^{-\alpha \phi}$ 
(where $\alpha$ is a non-negative constant) in the action (\ref{SD}) 
to obtain such solutions. 
The author would like to thank Nobuyoshi Ohta to point this out.}~\cite{Maeda:2009zi, Maeda:2009ds, Maeda:2010ja}:
\ba
H_{A_T} = \frac{t}{t_{A_T}} + \frac{Q_{A_T}}{r^{d-3}}\,.
\ea
Here 
$Q_{A_S}$ and $Q_{A_T}$ are brane charges.
$t_{A_T}$ is constant and determines time dependence of each brane $A_T$.
Note that $r^2:=\sum_i x_i^2$, so then $H_A$ never depends on $y_\alpha$. 
This means that 
all the branes are winding or smeared in all the $y_\alpha$ directions.
Then we assume here that 
the $y_\alpha$ directions are compactified on a torus $T^{D-d}$
and only the $x_i$ directions remain noncompact.

Let us consider dimensional reduction of all the $y_\alpha$ directions. 
The $x_i$ directions have spherical symmetry, so the metric in Einstein frame is
\ba\label{metric}
ds_d^2 =  \prod_A H_A^{\frac{1}{d-2}} \left[
\prod_A H_A^{-1} dt^2 + dr^2 + r^2 d\Omega_{d-2}^2
\right]
\ea
where
\ba
\prod_A H_A &=&
\prod_{A_S=1}^{n_S} H_{A_S} \prod_{A_T=1}^{n_T} H_{A_T}
\nt
&=&
\frac{\prod_A Q_A}{r^{(d-3)(n_S+n_T)}}
\prod_{A_S} \left(\frac{r^{d-3}}{Q_{A_S}}+1\right)
\prod_{A_T} \left(\frac{t}{t_{A_T}}\frac{r^{d-3}}{Q_{A_T}}+1\right)
\nt
&=:&
\left(\frac{R}{\tlr}\right)^{(d-3)(n_S+n_T)}\,.
\ea
Here $n_S$ and $n_T$ are the numbers of species of static and time-dependent branes, respectively.
And we define
\ba
R^{(d-3)(n_S+n_T)} &=& \prod_{A_S} \left(\frac{r^{d-3}}{Q_{A_S}}+1\right)
\prod_{A_T} \left(\frac{t}{t_{A_T}}\frac{r^{d-3}}{Q_{A_T}}+1\right)
\,,
\nt
\tlr^{(d-3)(n_S+n_T)} &=& \frac{r^{(d-3)(n_S+n_T)}}{\prod_A Q_A}\,.
\ea

In the following discussions, for simplicity, we set
\ba \label{common}
Q_{A_S} = Q_S\,,\quad
Q_{A_T} = Q_T\,, \quad
t_{A_T} = t_0
\ea
for all $A_S$ and $A_T$.
Then we obtain the expressions
\ba
R^{(d-3)(n_S+n_T)} 
=
\left(\frac{r^{d-3}}{Q_S}+1\right)^{n_S}
\left(\frac{t}{t_0}\frac{r^{d-3}}{Q_T}+1\right)^{n_T}, \quad
\tlr^{d-3} = \frac{r^{d-3}}{Q}\,,
\ea
where we define $Q:= (Q_S^{n_S} Q_T^{n_T})^\frac{1}{n_S+n_T}$.

\subsection{Region at infinite distance}

Let us now comment on the asymptotic behavior at an infinite distance $r\to\infty$.
We can rewrite the metric (\ref{metric}) as 
\ba
ds^2_d = -{\bar\Xi}^{d-3}d{\bar t}^2 + \frac{a^2}{\bar\Xi} (dr^2 + r^2d\Omega_{d-2}^2) 
\ea
where we define
\ba
\bar\Xi := \left( 1+\frac{Q_S}{r^{d-3}} \right)^{-\frac{n_S}{d-2}} \left(1+\frac{t_0}{t}\frac{Q_T}{r^{d-3}} \right)^{-\frac{n_T}{d-2}}
\ea
and
\ba
\frac{\bar t}{\bar t_0} = \left(\frac{t}{t_0}\right)^{1-\frac{(d-3)n_T}{2(d-2)}}
,\quad
a = \left(\frac{t}{t_0}\right)^\frac{n_T}{2(d-2)}.
\ea
Then in the limit of $r\to \infty$, this metric becomes
\ba
ds_d^2 = -d\bar t^2 + a^2 \left(dr^2+r^2d\Omega_{d-2}^2 \right).
\ea
This means that the time coordinate $\bar t$ is the proper time at infinity
and the metric in infinity is asymptotically Friedmann-Lema\^i{}tre-Robertson-Walker (FLRW) metric
with the scale factor $a\propto \bar t^{\,\frac{n_T}{2(d-2)-(d-3)n_T}}$.

Therefore, for $0<n_T\leq \frac{2(d-2)}{d-3}$, we have expanding FLRW universe. 
For $n_T>2$, we have accelerating universe (if $d>3$).
Especially in the case of $n_T=\frac{2(d-2)}{d-3}$, the scale factor is divergent.
This can be understood to describe the exponential expansion, i.e. de Sitter spacetime.
Later we will concentrate on the cases of $n_T + n_S = \frac{2(d-2)}{d-3}$, 
so there this de Sitter case is equivalently the $n_S=0$ case, where 
all the branes are time-dependent~\cite{Kastor:1992nn}.

\subsection{Near horizon region}

Now we look at the near horizon region  $r^{d-3}/Q_S \ll 1, r^{d-3}/Q_T \ll 1$.
If we also impose the condition $tr^{d-3}/t_0 Q_T \ll 1$, the metric (\ref{metric}) becomes 
\ba
ds_d^2 = \left(\frac{Q}{r^{d-3}}\right)^\frac{n_S+n_T}{d-2}
\left[
\left(\frac{r^{d-3}}{Q}\right)^{n_S+n_T} dt^2
+ dr^2 + r^2 d\Omega_{d-2}^2
\right]\,,
\ea
and the discussion becomes 
parallel to the time-independent cases $n_T=0$~\cite{Morita:2015jga}.
Therefore, let us here concentrate on the case of 
\ba \label{nhlim0}
\frac{r^{d-3}}{Q_S} \ll 1\,,\quad
\frac{r^{d-3}}{Q_T} \ll 1\,,\quad
\frac{t}{t_0} \frac{r^{d-3}}{Q_T} \simeq 1
\ea
to keep nontrivial time dependence. 
This means that when we define the variables as 
\ba
\tlt := \frac{t}{t_0}\,,\quad
\tlr^{d-3} := \frac{r^{d-3}}{Q}\,,
\ea
the near horizon limit can be defined by~\cite{Maeda:2009ds, Maeda:2010ja}
\ba \label{nhlim}
\tlt \to \frac{\tlt}{\eps}\,,\quad
\tlr^{d-3} \to \tlr^{d-3} \eps\,,\quad
\eps \to 0\,.
\ea

In this limit, all the terms in the metric (\ref{metric}) remain finite, only if 
\ba\label{nS}
n_T + n_S = \frac{2(d-2)}{d-3}\,.
\ea
Since $d$ and $n_T+n_S$ should be positive integers, 
all the combinations we need to consider are only 
\ba \label{cond-dn}
(d, n_T+n_S) = (1,1), (4,4), (5,3).
\ea
In the following discussions, we will concentrate on the latter two cases.\footnote{
In the $d=1$ case, all the spatial directions are compactified. At this moment we don't have any idea to discuss this case. 
}
Hereafter we fix $n_S$ as the relation (\ref{nS}) is satisfied and impose $n_T\neq 0$.

Then let us consider the metric in the near horizon limit.
In this limit we obtain 
\ba \label{R-def}
R^{(d-3)(n_T+n_S)} = (\tilde q \tlr^{d-3}+1)^{n_S} (q \tlt \tlr^{d-3}+1)^{n_T} 
~\to~  (q \tlt \tlr^{d-3}+1)^{n_T}
\ea
where $\tilde q:= Q/Q_S$ and $q:= Q/Q_T$, which are assumed to be of order one.
Using this relation, we can change the radial coordinate. 
The metric in this limit becomes
\ba
\frac{ds^2_\text{nh}}{Q^{\frac{2}{d-3}}} &=& 
-\frac{f(R)}{\tlt^2R^{2(d-3)}}d\tlt^2
-\frac{4(d-2)}{(d-3)^2n_T} \frac{1}{\tlt} \frac{R^{\frac{2(d-2)}{n_T}+1}}{R^{\frac{2(d-2)}{n_T}}-1} d\tlt dR
\nt
&&
+\frac{4(d-2)^2}{(d-3)^2n_T^2} \frac{R^{\frac{4(d-2)}{n_T}}}{\left(R^{\frac{2(d-2)}{n_T}}-1\right)^2} dR^2
+R^2 d\Omega_{d-2}^2
\ea
where we define
\ba
f(R):= \tau^2\left( R^{\frac{2(d-2)}{n_T}} -1 \right)^2
-\frac{R^{2(d-2)}}{(d-3)^2}\,,\quad
\tau = \frac{t_0}{q Q^{\frac{1}{d-3}}}\,.
\ea
Moreover, we can change the time coordinate so that the metric becomes a {\em static} form: 
\ba\label{met-st}
\frac{ds^2_\text{nh}}{Q^{\frac{2}{d-3}}}
= -\frac{f(R)}{R^{2(d-3)}} dT^2
+ \frac{4(d-2)^2}{(d-3)^2 n_T^2} \frac{\tau^2 R^{\frac{4(d-2)}{n_T}}}{f(R)} dR^2
+ R^2 d\Omega_{d-2}^2
\ea
where
\ba \label{T-def}
T = \pm \ln |\tlt| + \int^{R} \frac{2(d-2)}{(d-3)^2n_T}
\frac{R^{2d-5}}{\left(1-R^{-\frac{2(d-2)}{n_T}}\right)f(R)}dR\,.
\ea
This means that all the branes are static in this coordinate $(T, R)$.
Note that, however, this metric cannot be given in ordinary static black brane systems.
Thus we have obtain new examples to study the microstates of black branes 
in the $p$-soup model.

\section{Analysis based on p-soup model}
\label{sec3}

In the $p$-soup model, microstates of black branes are given by 
infinitely many (elementary) branes 
which are moving at the speed proportional to Hawking temperature 
and are gravitationally interacting with each other.
Therefore, in order to analyze black brane systems in this model, 
we need to write down effective action to describe the interactions among these branes.

Let us see details of the interactions.
First we choose one of the branes in the system as a probe
and consider the probe brane action 
\ba
S_{\text{probe,}A} = -\mu_A \left(
 \int d^{q_A+1}\xi \, \sqrt{-\det \gamma_{\mu\nu}} + \int \hat E_A
\right).
\ea
Here $\mu_A$ is the brane tension. 
$\hat E_A$ is the pullback of gauge potential to the brane worldvolume.
$\gamma_{\mu\nu}$ is the worldvolume metric induced from the spacetime metric (\ref{met-D}):
\ba
\gamma_{\mu\nu} = \partial_\mu Z^M \partial_\nu Z^N g_{MN}\, e^{-\frac{\epsilon_A a_A}{q_A+1}\phi}
\ea
where $\epsilon_A=1, -1$ for branes with electric/magnetic charges, respectively.
Note that we set the background metric for the probe brane $g_{MN}$ to the original black brane solution (\ref{met-D}) itself.
This can be justified since black brane systems are composed of infinitely many branes.
Even after we remove one of them (as a probe) from such a system, its metric must be nearly unchanged.

In the $D$-dimensional metric (\ref{met-D}), the harmonic functions $H_{A_S}, H_{A_T}$
depend on only time and $x_i$. 
It means that when we calculate the worldvolume metric,
we can neglect time dependence of brane's behavior in the $y_\alpha$ directions.
As we saw in \S 2, all the branes are winding or smeared on the torus $T^{D-d}$ 
in these directions, 
so this assumption is justified.

Therefore, we can take the static gauge for the coordinates $\xi$ on the probe brane worldvolume.
In addition, we assume that motion of the branes depends only on time $t$.
That is, position of the probe brane in the target spacetime $Z^M=Z^M(t)$.

Using these settings and integrating over the torus $T^{q_A}$ 
which the probe brane is winding around, 
the probe brane action becomes
\ba \label{action-probe}
S_{\text{probe,}A} = -m_A \int dt\left[
 \frac{1}{H_A}
 \sqrt{1-\left(\frac{d\vec r}{dt}\right)^2 \prod_{A'} H_{A'}}
 -\left(\frac{1}{H_A} -1 \right)
\right].
\ea
In the noncompact $d$-dimensional spacetime, 
the probe brane behaves as a BPS particle with the mass $m_A= \mu_A V_A$ (no sum of $A$).
$V_A$ is volume of the torus $T^{q_A}$.
$\vec r$ is position of the probe brane in the $d$ dimensions.

\subsection{Probe brane action in static coordinates}

In the $p$-soup model, as we mentioned in Introduction, velocities of the branes are important parameters.
Naively, in the probe brane action (\ref{action-probe}) we can define the velocity $\vec{v} = d\vec{r}/dt$.
However, the background here is time-dependent,
so this velocity seems not suitable to describe behaviors of the branes.
In order to avoid such problems, 
let us 
move to the time-independent frame, that is, the $(T, R)$ coordinates (\ref{met-st}).

Using eqs.(\ref{R-def}) and (\ref{T-def}), we can evaluate the norm of the velocity $\vec v$ as 
\ba 
 \left(\frac{d\vec r}{dt}\right)^2
= \left(\frac{1}{\tau q} \frac{d\tlr}{d\tlt}\right)^2
\ea
and 
\ba \label{drdt}
\left|\frac{d\tlr}{d\tlt}\right|
=
-\frac{ \left(R^{\frac{2(d-2)}{{n_T}}}-1\right)^{\frac{1}{d-3}}}{(d-3)q^\frac{1}{d-3} \tlt^\frac{d-2}{d-3} }
 \left(1- \frac{(d-3)^2 f(R)}{R^{2(d-2)}} \frac{g(R)\frac{dR}{dT}}{1-g(R)\frac{dR}{dT}}\right)\,.
\ea
Here $\tilde r$ is, more precisely, a vector in the noncompact $d$ dimensions.
$R$ denotes position of the probe brane in the $R$ coordinate, and we define
\ba
g(R) := \frac{2(d-2)}{(d-3)^2 n_T}\frac{R^{2d-5}}{\left(1-R^{-\frac{2 (d-2)}{n_T}} \right)  f(R)}\,.
\ea
Note that the $\tlt$ dependence in eq.(\ref{drdt}) disappears 
when we consider the combination
\ba \label{root}
\left(\frac{d\vec r}{dt}\right)^2\prod_{A'} H_{A'} 
=
\frac{1}{\tau^2}
\frac{R^{2(d-2)}}{(d-3)^2 \left(R^{\frac{2 (d-2)}{n_T}}-1\right)^2}
\left(1- \frac{(d-3)^2 f(R)}{R^{2(d-2)}} \frac{g(R)\frac{dR}{dT}}{1-g(R)\frac{dR}{dT}}\right)^2.
\ea
Here $H_{A'} = H_{A'}\big|_R$, since in the action (\ref{action-probe})
we consider the integration on worldvolume of the probe brane.
On the measure in the integration, we obtain the expression
\ba
\int \frac{dt}{H_A}
= \int \frac{d\tlt}{H_A} t_0 
\ea
and
\ba
\frac{d\tlt}{H_{A_S}} &=& \frac{Q}{Q_S} \tlr^{d-3} d\tlt 
= \left(R^{\frac{2(d-2)}{n_T}}-1\right) \frac{Q_T}{Q_S} \frac{d\tlt}{\tlt}
\,,\nt
\frac{d\tlt}{H_{A_T}} &=& \frac{Q}{Q_T} \frac{\tlr^{d-3}}{R^{\frac{2(d-2)}{n_T}}} d\tlt 
= \left( 1- R^{-\frac{2(d-2)}{n_T}} \right) \frac{d\tlt}{\tlt}
\ea 
for a static brane $A_S$ and a time-dependent brane $A_T$, respectively.
Then using the relation (\ref{T-def}), or equivalently,
\ba
\frac{d\tlt}{\tlt} = dT - g(R) dR = \left( 1-g(R) \frac{dR}{dT} \right) dT\,,
\ea
we can successfully eliminated the time coordinate $\tlt$.
Now the probe brane action (\ref{action-probe}) can be written in the static coordinate $(T,R)$,
except the rest mass term $-m_A \int dt$.
This term doesn't affect brane's behavior, 
so we will neglect it in the following analysis.

To summarize, the probe brane action for a static brane $A_S$ is 
\ba \label{probe-S}
S_{\text{probe,}A_S} = -m_{A_S} 
\frac{Q_T}{Q_S} t_0 
\int dT
\left(1-g(R)\frac{dR}{dT}\right)
\left(R^\frac{2(d-2)}{n_T}-1\right)
\left( \sqrt{1-\frac{h(R)^2}{(d-3)^2 \tau^2}} -1 \right)
\nt
\ea
and that for a time-dependent brane $A_T$ is
\ba \label{probe-T}
S_{\text{probe,}A_T} = -m_{A_T} 
t_0 
\int dT
\left(1-g(R)\frac{dR}{dT}\right)
\left( 1- R^{-\frac{2(d-2)}{n_T}} \right)
\left( \sqrt{1-\frac{h(R)^2}{(d-3)^2 \tau^2}} -1 \right)
\nt
\ea
where we define
\ba
h(R) := 
\frac{R^{d-2}}{R^{\frac{2(d-2)}{n_T}}-1}
\left(1- \frac{(d-3)^2 f(R)}{R^{2(d-2)}} \frac{g(R)\frac{dR}{dT}}{1-g(R)\frac{dR}{dT}}\right).
\ea

Let us here pay attention to the dependence on gravity coupling in $d$ dimensions $\kappa^2_d$.
The relation to $Q_S$ and $Q_T$ are given by
\ba \label{Q-m}
Q_S = \frac{2 m_{A_S} N_{A_S}}{(d-3)\Omega_{d-2}} \kappa_d^2\,,\quad
Q_T = \frac{2 m_{A_T} N_{A_T}}{(d-3)\Omega_{d-2}} \kappa_d^2\,,
\ea
where $N_A$ is the number of $q_A$-branes 
and $\Omega_{d-2}$ is the volume of a unit $(d-2)$-sphere.
Therefore, the expansion for small $\tau^{-1}= q Q^\frac{1}{d-3}/t_0$ means the expansion for small $\kappa_d^\frac{2}{d-3}$.
Since the last factor in the actions (\ref{probe-S}) and (\ref{probe-T}) can be expanded as
\ba \label{expandh}
\sqrt{1-\frac{h(R)^2}{(d-3)^2\tau^2}}-1
= -\sum_{n=1}^\infty \frac{(2n-3)!!}{2^n n!} \frac{h(R)^{2n}}{(d-3)^{2n}} \tau^{-2n},
\ea
this expansion form can be regarded as an expansion for small $\kappa_d^\frac{4}{d-3}$. 

Finally we comment on the period of the $T$ direction.
Naively, it should be the inverse temperature $\beta = 1/\cT$.
By taking into account the normalization, we can accurately obtain the period of the $T$ direction
\ba \label{norm-temp}
\frac{1}{\cT_\text{ren}} = \frac{(d-3)n_T}{2(d-2)} \frac{R^{d-3-\frac{2(d-2)}{n_T}}}{\tau} \frac{1}{Q^\frac{1}{d-3}} \frac{1}{\cT}\,.
\ea
Here the first two factors come from the renormalization of Killing vector 
(\ref{renom}), 
and the next $Q$ factor comes from the normalization in the static metric (\ref{met-st}).
Let us note that the temperature is low in our system 
so that moving branes compose a bound state and 
we can use the background metric (\ref{met-D}) in the extremal limit.

\subsection{Effective action}

Now we discuss the details of interactions among branes
and write down effective action to describe them.
In the previous subsection, 
we studied the interaction between a probe brane and the background.
Using this information, 
we can investigate how each brane interacts with other branes in this system.

Let us look at the factor $\tau^{-2n}$ in the expansion form (\ref{expandh}).
This can be written as
\ba
\tau^{-2n}
= \left(\frac{q Q^\frac{1}{d-3}}{t_0}\right)^{2n} 
= \left(\frac{Q_S^{n_S} Q_T^{n_T-2}}{t_0^2}\right)^n
\propto \kappa_d^\frac{4n}{d-3}\,.
\ea
We have already fixed $n_S$, but we use it here.
From the dependence on $Q_S$ and $Q_T$, we can find that this factor describes an interaction 
among $n_S n$ static branes and $(n_T-2) n$ time-dependent branes. 
By taking into account a probe brane, 
we find both the probe brane actions (\ref{probe-S}) and (\ref{probe-T}) describe
\begin{itemize}
\item interactions among $n_S n$ static branes and $(n_T-2)n+1$ time-dependent branes
\end{itemize}
where $n=1,2,\ldots,\infty$.
The number of interacting branes are $(n_S+n_T-2)n +1 = \frac{2n}{d-3}+1$,
and since $G_d = \kappa_d^2/8\pi$, we find $\frac{2n}{d-3}$ gravitons are 
exchanged among these branes. 
Note that we consider only the $d=4, 5$ cases, as we saw in eq.(\ref{cond-dn}), 
so the numbers of branes and gravitons are integers.

When $n_T=1$, the number of interacting time-dependent branes 
becomes negative. 
This may mean that our picture cannot describe the $n_T=1$ case,
and the $p$-soup model is valid only for the $n_T\geq 2$ cases. 
However, as we will see, the final results are correct also for the $n_T=1$ case.

In order to describe the interactions among the branes, 
the position of each brane $\vec R$ in the probe brane actions, 
or the vector from the center of the black brane background, 
should be replaced by the relative position of arbitrary two branes
\ba
\vec R_i - \vec R_j =: \vec R_{ij}\,.
\ea
Hereafter, we use the indices $i,j, \ldots$ which denote each brane in the system.
When we need to distinguish between static and time-dependent branes,
we use the indices $s_a$ for each static brane 
and $t_b$ for each time-dependent brane. 

The effective action should be written as a sum of all the interactions 
in the probe brane actions (\ref{probe-S}), (\ref{probe-T}), 
then we can write it down as 
\ba\label{Seff}
S_\text{eff} 
= \int dT~t_0 
\sum_{n=1}^\infty L_n
\ea
where 
\ba \label{Ln}
L_n &\sim&  
\left(\frac{\kappa_d^2}{(d-3)^{d-2}\Omega_{d-2}}\right)^\frac{2n}{d-3}
\sum_{\{s_1,\ldots,s_{n_Sn}\}} \sum_{\{t_1,\ldots,t_{(n_T-2)n+1}\}}
\prod_{a=1}^{n_Sn} m_{s_a} 
\prod_{b=1}^{(n_T-2)n+1} m_{t_b} 
\nt &&\quad\times
\left( 
\prod_{a\neq 1} {\frak h}(\vR_{s_1 s_a}) \prod_{b} {\frak h}(\vR_{s_1 t_b}) 
+\prod_{a} {\frak h}(\vR_{s_a t_1}) \prod_{b\neq 1} {\frak h}(\vR_{t_1 t_b}) 
+ \cdots
\right).
\ea
Hereafter `$\sim$' means an equality up to numerical (especially rational) factors.
${\frak h}(\vR_{ij})$ describes the interaction between two branes 
out of the interacting $\frac{2n}{d-3}+1$ branes,
which is defined as 
\ba \label{frakh}
\frak{h}(\vR_{ij}) =
\left[ \left(1-g(R_{ij})\frac{dR_{ij}}{dT}\right)
R_{ij}^{\frac{2(d-2)}{n_T}} {\frak f}(R_{ij}) 
\right]^\frac{d-3}{2n}
h(R_{ij})^{d-3}
\ea
where $R_{ij} := |\vR_{ij}|$.
The `$\cdots{\!\!\,}$' term includes the interactions of all the other combinations of the branes.
Note that in eq.(\ref{frakh}) we define a function 
\ba \label{fR}
{\frak f}(R) 
= {\frak c}_0  + {\frak c}_1 \left(1-R^{-\frac{2(d-2)}{n_T}}\right)
 + {\frak c}_2 \left(1-R^{-\frac{2(d-2)}{n_T}}\right)^2,
\ea
which is necessary 
because the second last factors of the probe brane actions 
(\ref{probe-S}) and (\ref{probe-T}) are slightly different from each other.
The coefficients ${\frak c}_{0,1,2}$ may be determined by
the details of interactions among the static and time-dependent branes.
In order to do it, we should give up the conditions (\ref{common})
and see the interactions of each brane more in detail.
However, for the remaining analysis, we can get enough information 
on the interactions from the effective action (\ref{Seff}),
so we don't do it here. This would be an interesting future work.

Although such ambiguity is in this analysis,
the physical quantities can be estimated,
as we will see in the following subsections.
It is because $R$ is dimensionless and assumed to be of order one
in the region (\ref{nhlim0}) we concentrate on.

\subsection{Evaluation of horizon radius and temperature}

Let us now estimate the physical quantities of our brane systems 
using the $p$-soup model.
As we mentioned in Introduction, first
we need to set the characteristic scales of size and velocity in the systems:
\ba
\vec R_{ij} \sim R\,,\quad
\frac{d \vec R_{ij}}{dT} \sim \frac{dR}{dT}
\ea
for all the branes $i, j$. 
This setting simplifies the following calculations. 
Next we impose the condition for these characteristic scales:
\ba \label{Mats}
\frac{dR}{dT} \sim \pi {\cal T}_\text{ren} R\,,
\ea
which may mean that we look at Matsubara modes of brane's behaviors.
Finally we impose the strong coupling condition, or virial theorem, such that
\ba \label{virial}
L_1 \sim L_2 \sim \cdots \sim \sum_{n=1}^\infty L_n\,,
\ea
since in the $p$-soup model the branes are strongly gravitationally interacting.
The final condition (\ref{virial}) means that eq.(\ref{root}) should be of order one,
so we can rewrite it as
\ba \label{cond}
\frac{h(R)^2}{(d-3)^2\tau^2} 
&=& 
\frac{R^{2(d-2)}}{\tau^2\left(R^{\frac{2 (d-2)}{n_T}}-1\right)^2}
\left[ \frac{1}{(d-3)^2}-\frac{f(R) g(R)}{R^{2(d-2)}} \frac{dR}{dT} - \frac{f(R) g(R)^2}{R^{2(d-2)}}  \left(\frac{dR}{dT}\right)^2 + \ldots \right]^2
\nt 
&\sim& 1.
\ea
Here we expand the last factor of eq.(\ref{root}) in a series of $dR/dT$.
Note that the velocity $dR/dT \ll 1$, 
because our brane system is assumed to be at low temperature $\cT_\text{ren}\ll 1$.
Then the first term should satisfy
\ba \label{cond1}
\frac{1}{(d-3)^2\tau^2} \frac{R^{2(d-2)}}{\left(R^{\frac{2 (d-2)}{n_T}}-1\right)^2} \sim 1\,,
\ea
which evaluates the characteristic size $R$. 
As we saw in eq.(\ref{cond-dn}), only the $n_T\leq 4$ cases are considered here,
so we can solve this condition for all the cases:
\ba \label{radius}
R\sim 
\begin{cases}
\left(\frac{\sqrt{1+4(d-3)^2\tau^2}\pm 1}{2\tau}\right)^\frac{1}{d-2} & \text{for $n_T=1$} \\
\left(1\pm\frac{1}{(d-3)\tau}\right)^\frac{1}{d-2} & \text{for $n_T=2$} \\
\left(\frac{18^\frac13 (d-3)^\frac13\tau^\frac13 (\sqrt{81-12(d-3)^2\tau^2}\pm 9)^\frac13}{12^\frac13 (d-3)^\frac23\tau^\frac23 + (\sqrt{81-12(d-3)^2\tau^2}\pm 9)^\frac23}\right)^\frac{3}{d-2} &\text{for $n_T=3$} \\
\left(\frac{(d-3)\tau}2 (1\pm \sqrt{1-\frac{4}{(d-3)\tau}})\right)^\frac{2}{d-2}, \quad
\left(\frac{(d-3)\tau}2 (\sqrt{1+\frac{4}{(d-3)\tau}}-1)\right)^\frac{2}{d-2} & \text{for $n_T=4$} \\
\end{cases}
\nt
\ea
where the double signs correspond in the $n_T=3$ case.
This result reproduces the horizon radius of corresponding time-dependent black brane system,
which can be easily checked: 
the horizon is at $f(R)=0$,
where the two sides of eq.(\ref{cond1}) become equal. 
This means that we correctly reproduce a result from supergravity.

The remaining terms of eq.(\ref{cond}) are proportional to
\ba
\frac{f(R)}{R^{2(d-2)}}\left(g(R)\frac{dR}{dT}\right)^n
\ea
for $n=1,2,\cdots$. 
Now we know that we are looking at the branes which are slowly moving $dR/dT\ll 1$ 
at the near horizon region $f(R) \sim 0$, $g(R)\gg 1$.
Here it seems natural to impose the condition 
\ba \label{cond3}
\frac{R^{2(d-2)}}{f(R)} \sim g(R)\frac{dR}{dT} \sim 1\,.
\ea
This means that all the terms in eq.(\ref{cond}) are of order one.
Then the second term should satisfy 
\ba \label{cond2}
\frac{f(R)g(R)}{R^{2(d-2)}}\frac{dR}{dT} 
= \frac{2(d-2)}{(d-3)^2n_T} \frac{1}{1-R^{-\frac{2(d-2)}{n_T}}} \frac{1}{R} \frac{dR}{dT}
\sim 1\,.
\ea
Using the setting (\ref{Mats}), 
the temperature of our system can be evaluated as 
\ba
\cT \sim
\frac{n_T^2}{\pi}
\left(1-R^{-\frac{2(d-2)}{n_T}}\right)
\frac{R^{d-3-\frac{2(d-2)}{n_T}}}{\tau Q^\frac{1}{d-3}}
\sim
\frac{n_T^2}{\pi R Q^\frac{1}{d-3}}\left(1-R^{-\frac{2(d-2)}{n_T}}\right)^2.
\ea
This is consistent with the result from supergravity (\ref{temp-sugra})
up to an overall rational factor and coefficient of each term.
In other words, we can reproduce the supergravity result up to the factor 
$\frac{{\frak a}_0 - {\frak a}_1 R^{-\frac{2(d-2)}{n_T}}}{{\frak a}_2 - {\frak a}_3 R^{-\frac{2(d-2)}{n_T}}}$,
where ${\frak a}_{0,1,2,3}$ are arbitrary rational numbers.

In the cases of static black branes, as we showed in our previous papers~\cite{Morita:2013wfa,Morita:2015jga},
we have only uncertainty of overall rational factors.
In the time-dependent black branes, on the other hand, we have another uncertainty
of factors including $R$.
However, this factor changes only the coefficients of $R^{-\frac{2(d-2)}{n_T}}$ in evaluated quantities, 
and $R$ is dimensionless and of order one.
Therefore, 
we can still claim that 
the results of order estimation in the time-dependent black branes 
are consistent with the supergravity results.

Finally note again that the temperature should be low in our brane system.
More precisely, our system should be 
in the near extremal limit $\cT \ll 1/RQ^\frac{1}{d-3}$ in supergravity.
This condition means $R\simeq 1$, and 
it is consistent with the near horizon limit (\ref{nhlim0}).

To summarize, 
the $p$-soup model can correctly tell us about 
the horizon radius and temperature of the time-dependent black branes.
In this picture, the branes are slowly moving at the near horizon region,
which ensures that the system is at low temperature.




\subsection{Evaluation of free energy and entropy}

Let us continue to estimate the physical quantities of our brane system.
The effective action (\ref{Seff}) is evaluated as
\ba
S_\text{eff} \sim \frac{t_0}{\cT_\text{ren}} L_1,
\ea
where we use the strong coupling condition (\ref{virial}).
Then the partition function can be estimated as $Z\sim e^{-S_\text{eff}}$,
and the free energy is defined as $\cF = -\cT\log Z$.
Therefore, the free energy of our system can be evaluated as
\ba
\cF &\sim& 
\frac{R^{d-3-\frac{2(d-2)}{n_T}} t_0}{\tau Q^\frac{1}{d-3}} L_1 
\sim
\frac{\Omega_{d-2}}{\kappa_d^2} Q R^{d-3} {\frak f}(R)
\nt
&=& 
\frac{\Omega_{d-2}}{\kappa_d^2} Q R^{d-3}
\left[ {\frak c}_0
+ {\frak c}_1 \left(1-R^{-\frac{2(d-2)}{n_T}}\right)
+ {\frak c}_2 \left(1-R^{-\frac{2(d-2)}{n_T}}\right)^2
\right],
\ea
where we use the conditions (\ref{cond1}) and (\ref{cond3}). 
This is perfectly consistent with the result from supergravity (\ref{F-sugra}).
The coefficients ${\frak c}_{0,1,2}$ cannot be determined in this analysis,
but we find here that they should depend on only the parameters $d$ and $n_T$.
This would be discussed in a future work.

Finally the entropy of our system is evaluated as
\ba
\cS = -\frac{\partial \cF}{\partial \cT}
\sim 
\frac{\pi \Omega_{d-2}}{\kappa_d^2}  Q^\frac{d-2}{d-3} R^{d-2}\,.
\ea
In the most right-hand side, the fractional expression 
where the numerator and denominator are polynomials in $R$ of the same degree
is set to one.
This can be justified, since 
our analysis has uncertainty of the factors including $R$,
as we discussed in the previous subsection.
Such a factor is included in this uncertainty. 
Therefore, we can correctly reproduce Bekenstein-Hawking entropy (\ref{BH-formula}).

In this way, we can show that the $p$-soup model can explain 
various thermodynamic quantities of the time-dependent black branes. 



\section{Conclusion and discussions}
\label{sec4}

In this paper, we discuss the $p$-soup proposal for a class of time-dependent black branes. 
Although they have many different properties from static black branes,
we can analyze them in a very similar way.
This may be partly because we can choose the time-independent frame (\ref{met-st}),
but their metrics in this frame are completely different from those of static black branes,
so this is undoubtedly a new nontrivial application of the $p$-soup model.

As a result, we find that 
the bound states of (elementary) branes in these systems exhibit 
the thermodynamic properties of the 
corresponding time-dependent black branes. 
This means that the $p$-soup analysis is applicable also for these systems
and that we get another evidence that 
the $p$-soup model describes (at least a part of) the microstates of a large class of black holes.


However, compared with the cases of static black branes,
our analysis holds 
subtleties.
For example, in the $n_T=1$ case, the $p$-soup picture of interacting branes 
seems not to be valid.
In all the cases there is some ambiguity about the factors including $R$.
The latter uncertainty is closely related to the undetermined coefficients in eq.(\ref{fR}), 
so we should analyze it more in detail and construct more plausible discussions in a future work.

Finally let us comment on the class of time-dependent black branes.
In the asymptotic region, the noncompact spacetime becomes FLRW universe.
Therefore, we can expect that this system is applied to some discussions in cosmology.
In particular, when all the branes are time-dependent (i.e.~$n_S=0$),
this universe has exponential expansion like inflation.
When a part of the branes are time-dependent (i.e.~$n_S, n_T\neq 0$), 
we have the universe with power law expansion.
Such properties may help us to draw up a scenario of making our own universe from branes.
Then, based on the $p$-soup model, it would be also interesting future works 
to discuss the systems of interacting branes creating various types of universe.




\subsection*{Acknowledgments}

The author would like to thank Takeshi Morita for useful comments.
This work is partially supported by Grant-in-Aid for Scientific Research 
(No.\,16K17711) from Japan Society for the Promotion of Science (JSPS).

\appendix

\section{Results from supergravity}
\label{secA}

The $d$-dimensional metric in Einstein frame (\ref{metric}) can be written as
\ba \label{met-sugra}
ds_d^2 = -\Xi^{d-3} dt^2 + \Xi^{-1} (dr^2 + r^2 d\Omega_{d-2}^2)
\ea
where
\ba
\Xi = \left(H_T^{n_T} H_S^{n_S}\right)^{-\frac{1}{d-2}}\,,\quad
H_T = \frac{t}{t_0} + \frac{Q_T}{r^{d-3}}\,,\quad
H_S = 1 + \frac{Q_S}{r^{d-3}}\,.
\ea
In the near horizon limit (\ref{nhlim}), it becomes
\ba
\frac{ds_\text{nh}^2}{Q^\frac{2}{d-3}} 
= - \frac{\tlr^{2(d-3)}}{(q\tlt \tlr^{d-3}+1)^{\frac{d-3}{d-2}n_T}}q^2\tau^2 d\tlt^2
+ \frac{(q\tlt \tlr^{d-3} +1)^\frac{n_T}{d-2}}{\tlr^{2}} (d\tlr^2+\tlr^2 d\Omega_{d-2}^2)\,.
\ea
This metric is invariant under the Killing vector $\xi^\mu$ defined by 
\ba
\xi^\mu := \tlt \left(\frac{\partial}{\partial\tlt}\right)^\mu - \tlr \left(\frac{\partial}{\partial\tlr}\right)^\mu.
\ea
In \S 2, we changed the coordinates and obtained the expression in the static form (\ref{met-st}).
In this coordinate $(T, R)$, 
this Killing vector is rewritten as $\xi^\mu = (\partial/\partial T)^\mu$.
Note that this vector becomes null at the horizon $f(R)=0$.

The surface gravity associated with the Killing vector $\xi^\mu$ is 
\ba\label{surface}
\kappa_\pm^2 = \mp \frac12 (\nabla_\mu \xi_\nu)(\nabla^\mu \xi^\nu)\,.
\ea
Then the surface gravities of horizons are evaluated as 
\ba \label{kappa}
\kappa_\pm 
= \pm\frac{(d-3)n_T}{4(d-2)} \frac{f'(R_\pm)}{\tau R_\pm^{d-3+\frac{2(d-2)}{n_T}}}
\ea
where $R=R_\pm$ are radii of the horizons.
As we saw in eq.(\ref{radius}), we have two event horizons in $n_T=1,2,3$ cases,
and three event horizons in $n_T=4$ case.
We can choose a suitable sign in eq.(\ref{kappa}) for each horizon.

However, when we calculate the surface gravity in a time-dependent spacetime, 
we need to care about the normalization of Killing vector.
In the case of a spherically symmetric spacetime, we should renormalize the Killing 
vector such that \cite{Maeda:2009ds}
\ba \label{renom}
\xi^\mu = \left(\frac{\partial}{\partial T}\right)^\mu ~\to~
\xi_\text{nh}^\mu  = \frac{(d-3)n_T}{2(d-2)} \frac{R^{d-3-\frac{2(d-2)}{n_T}}}{\tau} \left(\frac{\partial}{\partial T}\right)^\mu
\ea
where the renormalization factor is $(g_{TT}g_{RR})^{-\frac12}Q^\frac{2}{d-3}$ 
in the metric (\ref{met-st}).
Therefore, the surface gravities on the horizons should be evaluated 
using this Killing vector, and we obtain
\ba
\kappa_\text{nh}^\pm = \pm\frac{(d-3)^2 n_T^2}{8(d-2)^2} \frac{f'(R_\pm)}{\tau^2 R_\pm^{\frac{4(d-2)}{n_T}}}\,.
\ea
Using this surface gravity, the black hole temperature can be calculated as
\ba \label{temp-sugra}
T_\text{BH} := \frac{\kappa_\text{nh}^\pm}{2\pi Q^\frac{1}{d-3}}
= 
\frac{(d-3)^2 n_T^2}{8 (d-2)} \frac{1}{\pi R_\pm Q^\frac{1}{d-3}} \left(1-R_\pm^{-\frac{2(d-2)}{n_T}}\right)\left(1-\frac{2}{n_T}-R_\pm^{-\frac{2(d-2)}{n_T}}\right).
\ea
Here we have eliminated $\tau$ by the condition $f(R_\pm)=0$.

Next we discuss the black hole entropy.
This can be calculated using the Bekenstein-Hawking entropy formula.
Fortunately, the angular part of the metric (\ref{met-st}) is so simple, 
then we can easily obtain
\ba \label{BH-formula}
S_\text{BH} = \frac{\Omega_{d-2}}{4G_d} Q^\frac{d-2}{d-3} R_\pm^{d-2}
\ea
where $G_d =\kappa_d^2/8\pi$ 
is the Newton constant in $d$ dimensions.

Finally we discuss the free energy. 
It can be calculated as $F = E-TS$,
where the energy $E$ is ADM mass in ordinary supergravity calculations.
However, in our case the metric is globally time-dependent,
and it doesn't approach to a flat spacetime in asymptotic region.
This means there is no globally conserved energy. 

Instead, let us here discuss quasilocal energy
such as Misner-Sharp energy~\cite{Misner:1964je}.
Our metric (\ref{met-sugra}) has spherical symmetry, 
and we can define covariantly the circumference radius $\bar R:=r \Xi^{-\frac12}$, 
so the $d$-dimensional Misner-Sharp energy is given by~\cite{Maeda:2010ja, Maeda:2007uu}
\ba
E := \frac{d-2}{\kappa_d^2}\Omega_{d-2} {\bar R}^{d-3} \left[1-g^{\mu\nu}(\nabla_\mu \bar R)(\nabla_\nu \bar R)\right]\,.
\ea
Then in our case this energy is evaluated at the horizons as
\ba
E_\text{BH} = \frac{\Omega_{d-2}}{8\kappa_d^2}QR_\pm^{d-3}
\left[
4(d-2)-(d-4)n_T^2\left(1-R_\pm^{-\frac{2(d-2)}{n_T}}\right)^2\right]\,.
\ea
Therefore, the free energy can be obtained as
\ba \label{F-sugra}
F_\text{BH} = \frac{\Omega_{d-2}}{8(d-2)\kappa_d^2} Q R_\pm^{d-3}
\left[ c_0
+ c_1 \left(1-R_\pm^{-\frac{2(d-2)}{n_T}}\right)
+ c_2 \left(1-R_\pm^{-\frac{2(d-2)}{n_T}}\right)^2
\right]
\ea
where 
\ba
c_0 = 4 (d-2)^2\,,\quad
c_1 = 4 (d-3)^2 n_T \,,\quad
c_2 = -(3 d^2-18 d+26) n_T^2\,.
\ea



\begin{thebibliography}{99}

\bibitem{Strominger:1996sh}
  A.~Strominger and C.~Vafa,
  ``Microscopic origin of the Bekenstein-Hawking entropy,''
  Phys.\ Lett.\ B {\bf 379} (1996) 99
  [hep-th/9601029].

\bibitem{Callan:1996dv}
  C.~G.~Callan and J.~M.~Maldacena,
  ``D-brane approach to black hole quantum mechanics,''
  Nucl.\ Phys.\ B {\bf 472} (1996) 591
  [hep-th/9602043].

\bibitem{David:2002wn}
  J.~R.~David, G.~Mandal and S.~R.~Wadia,
  ``Microscopic formulation of black holes in string theory,''
  Phys.\ Rept.\  {\bf 369} (2002) 549
  [hep-th/0203048].

\bibitem{Wiseman:2013cda}
  T.~Wiseman,
  ``On black hole thermodynamics from super Yang-Mills,''
  JHEP {\bf 1307} (2013) 101
  [arXiv:1304.3938 [hep-th]].

\bibitem{Morita:2013wfa}
  T.~Morita, S.~Shiba, T.~Wiseman and B.~Withers,
  ``Warm $p$-soup and near extremal black holes,''
  Class.\ Quant.\ Grav.\  {\bf 31} (2014) 085001
  [arXiv:1311.6540 [hep-th]].

\bibitem{Morita:2014ypa}
  T.~Morita, S.~Shiba, T.~Wiseman and B.~Withers,
  ``Moduli dynamics as a predictive tool for thermal maximally supersymmetric Yang-Mills at large $N$,''
  JHEP {\bf 1507} (2015) 047
  [arXiv:1412.3939 [hep-th]].

\bibitem{Morita:2013wla}
  T.~Morita and S.~Shiba,
  ``Thermodynamics of black M-branes from SCFTs,''
  JHEP {\bf 1307} (2013) 100
  [arXiv:1305.0789 [hep-th]].

\bibitem{Morita:2014cfa}
  T.~Morita and S.~Shiba,
  ``Microstates of D1-D5(-P) black holes as interacting D-branes,''
  Phys.\ Lett.\ B {\bf 747} (2015) 164
  [arXiv:1410.8319 [hep-th]].

\bibitem{Morita:2015jga}
  T.~Morita and S.~Shiba,
  ``Thermodynamics of Intersecting Black Branes from Interacting Elementary Branes,''
  JHEP {\bf 1509} (2015) 070
  [arXiv:1507.03507 [hep-th]].

\bibitem{Maeda:2009zi}
  K.~Maeda, N.~Ohta and K.~Uzawa,
  ``Dynamics of intersecting brane systems -Classification and their applications-,''
  JHEP {\bf 0906} (2009) 051
  [arXiv:0903.5483 [hep-th]].

\bibitem{Maeda:2009ds}
  K.~Maeda and M.~Nozawa,
  ``Black Hole in the Expanding Universe from Intersecting Branes,''
  Phys.\ Rev.\ D {\bf 81} (2010) 044017
  [arXiv:0912.2811 [hep-th]].

\bibitem{Maeda:2010ja}
  K.~Maeda and M.~Nozawa,
  ``Black Hole in the Expanding Universe with Arbitrary Power-Law Expansion,''
  Phys.\ Rev.\ D {\bf 81} (2010) 124038
  [arXiv:1003.2849 [gr-qc]].

\bibitem{Kastor:1992nn}
  D.~Kastor and J.~H.~Traschen,
  ``Cosmological multi-black hole solutions,''
  Phys.\ Rev.\ D {\bf 47} (1993) 5370
  [hep-th/9212035].

\bibitem{Misner:1964je}
  C.~W.~Misner and D.~H.~Sharp,
  ``Relativistic equations for adiabatic, spherically symmetric gravitational collapse,''
  Phys.\ Rev.\  {\bf 136} (1964) B571.

\bibitem{Maeda:2007uu}
  H.~Maeda and M.~Nozawa,
  ``Generalized Misner-Sharp quasi-local mass in Einstein-Gauss-Bonnet gravity,''
  Phys.\ Rev.\ D {\bf 77} (2008) 064031
  [arXiv:0709.1199 [hep-th]].

\end{thebibliography}
\end{document}